%% file: 0Main_Paper.tex
\documentclass[twocolumn, 10pt]{IEEEtran}           

\input{0Config.tex}
\input{0Glossary.tex}
\allowdisplaybreaks
\begin{document}
	\title{Decentralized multi-agent reinforcement learning with shared actions}

	\author{Rajesh K Mishra, Deepanshu Vasal and Sriram Vishwanath\\
	University of Texas at Austin\\
	USA
	}

	\maketitle

	\begin{abstract}
		In this paper, we propose a novel model-free reinforcement learning algorithm to compute the optimal policies for a multi-agent system with $N$ cooperative agents where each agent privately observes it's own private type and publicly observes each others' actions. The goal is to maximize their collective reward. The problem belongs to the broad class of decentralized control problems with partial information. We use the common agent approach~\cite{Nayyar} wherein some fictitious common agent picks the best policy based on a belief on the current states of the agents. These beliefs are updated individually for each agent from their current belief and action histories. Belief state updates without the knowledge of system dynamics is a challenge. In this paper, we employ particle filters called the \emph{bootstrap} filter distributively across agents to update the belief. We provide a model-free \rl method for this multi-agent \pomdp using the particle filter and sampled trajectories to estimate the optimal policies for the agents. We showcase our results with the help of a smartgrid application where the users strive to reduce collective cost of power for all the agents in the grid. Finally, we compare the performances for model  and model-free implementation of the \rl algorithm establishing the effectiveness of \pf method.
	\end{abstract}

	\begin{IEEEkeywords}
		Decentralized control, Reinforcement Learning, Particle Filters
	\end{IEEEkeywords}

	\section{Introduction}
		\input{1Introduction.tex}
	\section{Model}
		\label{sec:Model}
		\input{2Model.tex}

	\section{Preliminaries}
		\label{sec:Prelims} 
		\input{3Preliminaries.tex}

	\section{Convergence}
		\label{sec:Convergence}
		\input{4Convergence.tex}
	\section{Reinforcement Learning Algorithm}
		\label{sec:Algo}
		\input{5Algorithm.tex}
	\section{Conclusion}
		\label{sec:concl}
		\input{6Conclusion.tex}
	\appendix
	\section*{Appendix A}
		\input{7Appendix.tex}

	\medskip	
	\small
	\bibliographystyle{IEEEtran}
	\bibliography{References}
\end{document}

%% file: 0Config.tex
\usepackage{bbm}
\usepackage{amsmath,amssymb,amsfonts}
\usepackage{mathrsfs}
\usepackage{xcolor}
\usepackage{graphicx}
\usepackage{amsfonts}
\usepackage[english]{babel}
\usepackage{amsthm}
\usepackage{epstopdf}
\usepackage{textcomp}
\usepackage[linesnumbered,ruled,vlined]{algorithm2e}
\usepackage{mathtools}

\newtheorem{theorem}{Theorem}

\newtheorem{lemma}{Lemma}

\newcommand{\bN}{[N]}

\newcommand{\nn}{\nonumber}

\newcommand{\defeq}{\buildrel\triangle\over =}
\newcommand{\sq}[1]{\left[#1\right]}
\newcommand{\cm}[1]{\left(#1\right)}
\newcommand{\bm}[1]{\left\{#1\right\}}

\newcommand{\figref}[1]{Fig.~{\ref{#1}}}

\newcommand{\mP}{\mathbb{P}}
\newcommand{\mE}{\mathbb{E}}
\newcommand{\mR}{\mathbb{R}}

\newcommand{\cX}{\mathcal{X}}

\newcommand{\cA}{\mathcal{A}}

\newcommand{\cH}{\mathcal{H}}
\newcommand{\cT}{\mathcal{T}}
\newcommand{\cN}{\mathcal{N}}

\newcommand{\tgamma}{\tilde{\gamma}}

\newcommand{\pibar}{\boldsymbol{\pi}}
\newcommand{\pitbar}{\boldsymbol{\hat{\pi}}}
\newcommand{\abar}{\boldsymbol{a}}
\newcommand{\Fbar}{\boldsymbol{F}}
\newcommand{\gbar}{\boldsymbol{\gamma}}
\newcommand{\gtbar}{\boldsymbol{\tgamma}}
\newcommand{\Xbar}{\boldsymbol{X}}
\newcommand{\xbar}{\boldsymbol{x}}
\newcommand{\Abar}{\boldsymbol{A}}

\newcommand{\Ftbar}{\boldsymbol{\hat{F}}}
\newcommand{\Tbar}{\boldsymbol{\tilde{\theta}}}
\newcommand{\Ttbar}{\boldsymbol{\hat{\theta}}}
\newcommand{\Tbold}{\boldsymbol{\theta}}

\newcommand{\thehat}{\hat{\theta}}
\newcommand{\pihat}{\hat{\pi}}
\newcommand{\Qhat}{\hat{Q}}
\newcommand{\Vhat}{\hat{V}}

\newcommand{\Thetahat}{\hat{\theta}}
\newcommand{\Jhat}{\hat{J}}

\SetKwInput{KwInput}{Input}
\SetKwInput{KwOutput}{Output}

%% file: 0Glossary.tex
\usepackage{xspace}
\usepackage[acronym,nogroupskip,nonumberlist,nopostdot]{glossaries}

\newacronym{mdp}{MDP}{Markov decision process}
\newacronym{ne}{NE}{Nash equillibrium}
\newacronym{mfe}{MFE}{mean-field equillibrium}
\newacronym{mpe}{MPE}{Markov perfect equillibrium}
\newacronym{mfg}{MFG}{mean-field game}
\newacronym{rl}{RL}{reinforcement learning}
\newacronym{marl}{MARL}{multi-agent reinforcement learning}
\newacronym{iot}{IoT}{Internet of Things}
\newacronym{ssg}{SSG}{Stackelberg Security Game}
\newacronym{pse}{PSE}{Perfect Stackelberg Equilibrium}
\newacronym{mpse}{MPSE}{Markov \gls{pse}}
\newacronym{spbe}{SPBE}{Sub-game Perfect Bayesian Equilibrium}
\newacronym{pf}{PF}{particle filter}
\newacronym{pomdp}{POMDP}{partially obseravable \gls{mdp}}
\newacronym{nn}{NN}{neural network}

\newcommand{\mdp}{\gls{mdp}\xspace}

\newcommand{\rl}{\gls{rl}\xspace}
\newcommand{\marl}{\gls{marl}\xspace}

\newcommand{\pf}{\gls{pf}\xspace}
\newcommand{\pomdp}{\glspl{pomdp}\xspace}
\newcommand{\neural}{\gls{nn}\xspace}

%% file: 1Introduction.tex
Multi-agent systems are ubiquitous. Communication networks, autonomous cars, traffic management systems, power grids~\cite{bard1902hanabi,stone1999team,fax2004information,schneider1999distributed}, are just few of the countless examples of such systems in our daily lives. Several agents, interacting with the same environment, with actions affecting both the environment and other agents in the system, create interdependencies which have made the study of such systems challenging and at times intractable. Centralized multi-agent systems have a central agent that can have access to all the available information and takes a unified action optimized globally. On the other hand, systems with decentralized control have many agents possessing different information about the same environment. In this paper, we are considering a multi-agent system with decentralized control. In addition, the agents act in a cooperative manner to maximize the collective reward rather than their sole self-interest at the behest of other agents. 

We consider a specific decentralized cooperative team problem with incomplete information where the model of the system is unknown. Usually, the systems with partial observation of states are handled by considering a belief which is the probability of the system being in certain state given the observed behavior. In the model-free version, we use sequential monte-carlo methods like the particle filters to estimate the belief state. We develop a backward recursion algorithm to solve for the optimal strategy for such problems through a common agent~\cite{Nayyar} which computes the belief on the private state information of all the agents and provides the optimal action. We proceed to implement a reinforcement learning algorithm using sampled trajectories for an example numerical problem. We implement the above described techniques in a smartgrid application, where we try to minimize the cost associated with the grid participants taking actions to modulate their power consumption. We plot the results and show the comparison both when the model is known and when the model is not known. We also provide the proof of convergence of the above techniques.

\subsection{Related Work}
Authors in ~\cite{Asghari2020} consider a system that is similar to the set up that was considered in this paper. It has a multi-agent system set up with decentralized control and cooperative agents. They consider an LQ system with agents with different degrees of communication between the agents. They do a regret analysis for all these cases both for single and multi-agent systems with respect to the time horizon. In ~\cite{Nayyar2011}, the authors consider decentralized control systems and investigate $n$-step delayed information structure comprising $K$ controllers. They also show that that the idea of formulating an equivalent problem from the point of view of a coordinator that has access to all the information which is common to all agents is useful for decentralized control problems. It could be considered as a version of the problem we tackle in the paper but with the model known. Authors in~\cite{Subramanian2019} put forth the concept of an approximate information state capturing the idea of the belief state in a partially observable mdp. They show that the information state is a sufficient statistic in order to obtain the dynamic program that solves the \pomdp. They also propose an \rl construction that achieves this with an approximate information state and could be obtained by sampling trajectories from the system. In~\cite{Zhang2018}, the authors consider a communication network as a decentralized multi-agent problem. They provide a actor critic based \rl algorithm that solves the problem when the agents locally observe a state and the communication over the agents gives additional information to the agents along with convergence results. Authors of the paper~\cite{Gupta2017} talk about deep \marl and feed forward neural architectures. They consider cooperative policies in a partially observable domain with no communication between the agents. They do a comparative study on TD algorithms actor critic and on policy gradient approaches and show that the policy gradient method outperforms.~\cite{Omidshafiei2017} formalizes and addresses the problem of multi-task multi-agent reinforcement learning under partial observability.

The structure of the paper is as follows.Section~\ref{sec:Model} presents the decentralized \pomdp model followed by the Section~\ref{sec:Prelims} that describe the theory and the various concepts used in the paper including the common agent approach and the particle filter algorithm. Section~\ref{sec:Convergence} discusses the convergence of the algorithms with the particle filters to the same expected returns as the case when the model is known. Section~\ref{sec:Algo} gives the reinforcement learning implementation of the above method for a smart grid application and conclude it with results and conclusions. 

\subsection{Notation}

We use uppercase letters for random variables and lowercase for their realizations. Similarly, letters with normal fonts represent scalars where as boldfaced letters represent a vector of the same variables. Subscripts on variables, represent time indices while superscripts represent the agent identities. For any finite set $\mathcal{S}$, $\Delta\mathcal{S}$ represents space of probability measures on $\mathcal{S}$ and $|\mathcal{S}|$ represents its cardinality. We denote by $P^{\sigma}$ (or $E^{\sigma}$) the probability measure generated by (or expectation with respect to) strategy profile $\sigma$. We denote the set of real numbers by $\mathbb{R}$. All equalities and inequalities involving random variables are to be interpreted in \emph{a.s.} sense. 

%% file: 2Model.tex
Consider a discrete multi-agent system comprising $N$ cooperative agents with $\cN$ representing the set of agents. Let $\cm{\cX^i}_{i\in\cN}$ and $\cm{\cA^i}_{i\in\cN}$ be their state and action spaces respectively. At time $t$, any agent $i\in\cN$ can completely observe its own private state $x_t^i\in\cX^i$ along with the action histories $\abar_{1:t-1}\coloneqq\cm{a^i_{1:t-1}}_{i\in\cN}$ of all agents. However, it is oblivious to the states of all the other agents in the system. With the knowledge of $x_{1:t-1}^i$ and $\abar_{1:t-1}$, the agent can take an action $a_t^i\sim\sigma_t^i\cm{\cdot\vert x_{1:t-1}^i,\abar_{1:t-1}}$ where $\sigma^i_t$ be some probabilistic strategy of agent $i$ defined as $\sigma^i_t : \cA^{t-1}\times (\cX^{i})^t \to \cA^i$. The state of the individual agent then evolves as a conditionally independent Markov process given as
\begin{align}
    \mP\cm{x_t^i\vert x_{t-1}^i,a_t^{i-1}} = \tau_t^i\cm{x_t^i\vert \xbar_{t-1},\abar_{t-1}},
\end{align}
where $\tau^i$ represents the dynamics of the model with respect to agent $i$. The state of the entire system $\xbar_t\coloneqq\cm{x^i_t}_{i\in\cN}$ then evolves as 
\begin{align}
    \mP(\xbar_t \vert \xbar_{t-1},\abar_{t-1}) &= \prod_{i=1}^N \tau_t^i\cm{x_t^i\vert \xbar_{t-1},\abar_{t-1}}.
\end{align}  
Consequently, the agents receive a common stationary reward $R\cm{\xbar_t,\abar_t}$ as a function of the state and action vectors corresponding to the agents of the entire system. 

In this paper, we consider a model-free version of the problem and so assume that the dynamics are unknown to the agents. The model is viewed as a blackbox and is only limited to the following two uses:
\begin{enumerate}
    \item The model can be initialized with some some state distribution $\xbar_0\sim \boldsymbol{p}\cm{\cdot}$, based on the policy chooses an action and undergoes a state transition.
    \item The model can also be employed to generate a series of state transitions given certain current states and actions through sampling to be used as a part of the proposed filter.
\end{enumerate}
However, we do consider the version with the model known as a way to compare and benchmark the results from model-free versions.

We implement the algorithms by considering both finite time horizon case, where the model behavior is a function of $t\in\cT\coloneqq\bm{1\ldots T}$, and the infinite time horizon case, where the dynamics are time-invariant. The algorithms aim to obtain the optimal policies for the agents which maximizes the expected sum of the discounted rewards from the model given as 
\begin{align}
    J_t = \mE\sq{\sum_{n=t}^T\delta^{n-t}R\cm{\xbar_n,\abar_n}\vert \boldsymbol{\cH_{t-1}}},
\end{align}
where $ \boldsymbol{\cH_{t-1}}=\bm{\abar_{1:t-1}}$ represents the past history till time $t-1$, and the expectation is over all randomness in the system. $\delta$ is the discount factor. For the infinite horizon case, the definition of $J_t$ remains the same with $T$ being replaced with $\infty$.

%% file: 3Preliminaries.tex
The decentralized control system that we consider, has the agents act like a team to maximize their collective reward, being privy to the local state information alone. Such a system can be modeled as a dynamic team with asymmetric information, denoted as $\mathfrak{D}_T$ for finite horizon and $ \mathfrak{D}_\infty $ for infinite horizon. In our paper, we consider the common agent approach to solve for the optimal policy for the agents. This approach was proposed by the authors in~\cite{Nayyar} where the decentralized problem was reformulated as an equivalent centralized problem with a coordinator. The coordinator knows the common information and then proposes prescriptions that map the local information of the agents to their control action. In a similar approach, we consider an arbitrary common agent computes a belief on the states of the agents based on the observed actions and provides a policy, optimized over the belief, that would maximize the collective reward.

\subsection{Common agent approach}\label{sec:common_agent}

  Let us consider the finite horizon dynamic system model $\mathfrak{D}_{T}$. Consider a fictitious common agent which only observes the action histories of the agents at time $t$ and computes a belief on the states defined as $\pi_t^i=\mP^{\boldsymbol{\sigma}}\cm{x_t^i\vert\abar_{1:t-1}}\ \forall i\in\cN$. For any strategy of the agents, the joint common belief can be factorized as a product of its marginals i.e., $\pibar_t(x_t) = \prod_{i=1}^N \pi_t^{i}(x_t^i), \forall x_t$. In this paper, we only deal with joint beliefs and to accentuate this independence structure, we define $\pibar_t \in \times_{i\in \cN} \Delta(\cX^i)$ as vector of marginal beliefs where $\pibar_t := (\pi^i_t)_{i\in \cN}$. We also define an action policy $\gamma_t^i$ for the underlying agent $i$ to be $\gamma_t^i\colon\cX^i\to\cA^i$ which is a pure policy specifying the action that the agent can take given a particular private state. The common agent computes the global optimal policy $\Tbar=\cm{\tilde{\theta}^i_t}_{i\in\cN,t\in\cT}$, where $\tilde{\theta}^i_t : \times_{i\in\cN} \Delta\cm{\cX^i} \to \bm{\cX^i \to \cA^i }$, which maps the belief state vector $\pitbar_t$ to the optimal policy vector $\gtbar\coloneqq\cm{\tgamma_t^i}_{i\in\cN}$ corresponding to all agents. When the common agent, for some current belief $\pibar_t$ plays the policy $\gbar_t$ and observes the actions $\abar_t$ from the agents, we can define the vector of belief updates as $\Fbar(\pibar,\gbar,\abar) := (F^i(\pi^i,\gamma^i,a^i))_{i \in \cN}$ where (using Bayes rule)
  \begin{align}
    \label{Eqn:Bayes}
    &F^i\cm{\pi_t^i,\gamma_t^i,a_t^i}\cm{x^i_{t+1}}= 
    &\left\{
    \begin{array}{ll}
      \frac{N_{x_t^i}}{D_{\tilde{x}_t^i}}\quad\text{ if }D_{\tilde{x}_t^i}\neq0\\
      N^\prime_{x_t^i}
    \end{array}
    \right.
  \end{align}
  with 
  \begin{align}
    N_{x_t^i}         &=  \sum_{x^i_t} \pi_t^{{i}}(x_t^i) \gamma_t^i(a_t^i|x_t^i)\tau_t^i(x_{t+1}^i|x_t^i, a_t^i)\\
    N^\prime_{x_t^i}  &=  \sum_{x_t^i}\pi_t^i(x_t^i)\tau_t^i(x_{t+1}^i|x_t^i,a_t^i)\\
    D_{\tilde{x}_t^i} &=  \sum_{\tilde{x}_t^i}\pi_t^{{i}}(\tilde{x}_t^i)  \gamma_t^i(a_t^i|\tilde{x}_t^i).
  \end{align}
  
  The update function $ F^i $ is a function of time $ t $ through the kernel $\tau_t^i$ (for the finite horizon model). However, for notational simplicity we suppress this dependence on $t$.
  
\subsection{Particle Filter}
  
  In this section, we discuss a model-free method to update the belief function vector in~\eqref{Eqn:Bayes}. The main challenge in estimating the updated beliefs as per the Bayesian update is the inability to evaluate it without the explicit knowledge of the transition functions $\tau_t^i$'s. Here we discuss a sampling based method to compute the required solution.  
  
  Kalman filters have been widely used for Gaussian state space modelling but the advent of sequential monte carlo methods can be attributed to certain applications where non-Gaussian state space modelling was required~\cite{kitagawa1996monte}. Particle filters are sequential monte carlo filters that approximate the belief state from an empirical distribution based on the observed history when model dynamics are unknown~\cite{Douc2005,Crisan2002}. It is widely popular in applications like robotics for localizations and fault dynamics, where most of the times the environment is non-Gaussian and needs to learned based on collected samples and observations~\cite{rekleitis2004particle}. These methods utilize a $K$ number of random samples or \emph{particles}, where $K$ is large, to represent the posterior probability of the state based on the observations.

  The particles are propagated over time using sequential importance sampling and resampling steps. The resampling step statistically multiplies and/or discards particles at each time step to adaptively concentrate particles in regions of high posterior probability based on the observation. These methods are very flexible and can be easily applied to nonlinear and non-Gaussian dynamic models. Such methods have been used in many applications~\cite{Douc2005}\cite{Crisan2002}. \pf approximates the belief state $\pi_t^i,i\in\bN$ by a set of $K$ sampled points from the state space $x_t^i\in\cX$, updated in a sequential manner at every observation point $a_t^i$, which also serves as an action in our case, through a selection procedure to establish the truthfulness of the belief based on the observation. In other words, the belief using a \pf could be expressed as,

  \begin{align}
    \pihat=\frac{1}{K}\sum_{i=1}^Kf\cm{x_t^i}.
  \end{align}
  The generic \pf called the bootstrap filter that samples the states from a previous distribution and the resamples based on observations. It is estimated using an empirical distribution given as 
  \begin{align}
    \pihat^{i}\cm{x_t^i} \defeq \sum_{j=1}^K\delta_{x^i_t}w_t^j,
  \end{align}
  where $\delta_{x^i_t}$ is a dirac delta function made up of $K$ particles $x_t^{i,1:K}$. The algorithm recursively consists of two steps, a transition step to sample $K$ particles from the current distribution and to obtain the samples corresponding to the next states for each of the sample according to the transition function. This is followed by a selection step where it is resampled according to the weights $w_t^i$ generated based on the observations. The algorithm can be summarized as below~\cite{Elvira2017,Doshi2009,Coquelin}.

  \begin{enumerate}
    \item Initialize, $t = 0$, $x_0^i\sim\pi_t\cm{\cdot}$, set $t=1$,
    \item For $t = 1,2,3,\ldots$,\newline 
    \textbf{Importance sampling:}
    \begin{enumerate}
      \item For $ i =1 \ldots K$, sample from the model, $\hat{x}_i\sim\mP\cm{x_{t+1}\vert x_t,a_t}$
      \item For $ i = 1\ldots K$, compute the weights in proportion to the chances of the next state with the current observation $a_t$
      \begin{align}
        w_t^i=\frac{\mP^\theta_i\cm{a_t,\hat{x}_i}}{\sum_{j=1}^K\mP^\theta_i\cm{a_t,\hat{x}_j}}
      \end{align}
    \end{enumerate}
    \textbf{Selection/ resampling}
    \begin{enumerate}
      \item Resample from the list $\cm{\hat{x}_{t+1,i}}$ with replacement according to the weights to get $\cm{x_{t+1,i}}$. This is done by choosing from indices $\bm{1\ldots K}$ according to the multinomial distribution $\cm{w_1\ldots w_K}$.
      \item The new belief state estimate is then,
      \begin{align}
        \pihat^{i,K}\cm{x_t^i} \defeq \sum_{i=1}^K\delta_{x^i_{t+1}}
      \end{align}
    \end{enumerate}
  \end{enumerate}
  
  The multinomial distribution used for resampling is one of the simplest methods that was introduced in~\cite{Vapnik1999}. This methods redistributes the samples based on their corresponding weights. Other versions of the resampling method include the stratified sampling method that reduces variance~\cite{kitagawa1996monte,Douc2005}.

  The common agent employs a series of $N$ parallel particle filters that estimate $\mP^\theta_i\cm{x^i_{t+1}\vert\abar_{1:T}}$ for each of the agent given the policy $\gbar_t$. The particle filter as a module takes in the current belief vector, the corresponding policy and the observation vector. It uses the model to sample the next steps and then computes the posterior distribution.

  \subsection{Backward Recursion} \label{sec:fhbr}
  
  We provide a backward recursion algorithm for $\mathfrak{D}_T$ to compute the optimum policy $\Tbar$ for the common agent which was discussed in Section~\ref{sec:common_agent}. Though, this algorithm is an inefficient way to compute the optimal policy, it is stated as a reference and used for proving convergence results later~\cite{Subramanian2019}. This algorithm was also used to compute the optimal returns that could be achieved out of the system and forms the baseline to compare the performance of our \rl algorithm. 
  
  We define a sequence of action-value function `$Q_t$' and value function `$V_t$' for agents at time $t$ defined as
  \begin{align}
    &Q_t:\cm{\times_{i\in\cN} \Delta(\cX^i)}\times\cm{\times_{i\in\cN}\bm{\cX^i \to \cA^i}}\to \mR\nn\\
    &V_t : \cm{\times_{i\in\cN} \Delta(\cX^i)}\to \mR.\nn
  \end{align} 
  The bellman update for the action value function $Q_t$ can be expressed as
  \begin{align}
    \label{Eqn:Bellman_Q_update}
    Q_t\cm{\pibar_t,\gbar_t} \defeq \mE^{\pibar_t,\gbar_t}&\left[R\cm{\Xbar_t,\Abar_t}+\right.\nn\\
    &\left.\delta V_{t+1}\cm{\Fbar\cm{\pibar_t, \gbar_t, \Abar_t}}\right],
  \end{align}
  where the expectation is over the random variables $\cm{\Xbar_t,\Abar_t}$ with the distribution $\gbar_t\cm{\cdot\vert\xbar_t}\pibar_t\cm{\xbar_t}$. The value function at certain belief $\pibar_t$ is computed from the policy $\gtbar_t$ that maximizes the $Q_t$ at the same $\pibar_t$. This can be summarized as 
  \begin{align}
    V_t\cm{\pibar_t} = Q_t\cm{\pibar_t, \gtbar_t}, \label{Eqn:Bellman_V_update1}
  \end{align}
  where
  \begin{align}  
    \gtbar_t \in \arg\max_{\gbar_t} \ Q_t\cm{\pibar_t,\gbar_t}\label{Eqn:Bellman_V_update2}.
  \end{align}
  Now, let us replace the true belief $\pibar_t$ with the estimated distribution $\pitbar_t$ from the particle filters. We define different action-value functions $\Qhat_t$ and value functions $\Vhat_t$ in the same way as before. These quantities are generated through a backward recursive way, as follows:
  
  \begin{enumerate}
    \item Initialize $\forall \pitbar_{T+1}\in \times_{i\in\bN} \Delta(\cX^i)$,
    \begin{align}
      \Vhat_{T+1}\cm{\pitbar_{T+1}} \defeq 0 \label{Eqn:VT+1_1}.
    \end{align}     
    \item For $t = T,T-1, \ldots 1$,  
    \begin{enumerate}
      \item Compute $\Qhat \ \forall\pitbar_t\in \times_{i\in\bN} \Delta(\cX^i)$ and $\forall\gbar_t\in \times_{i\in\bN}\Gamma^i$ similar to the equation in~\eqref{Eqn:Bellman_Q_update}, as follows:
      \begin{align}
        \label{Eqn:Qdef1}
        \Qhat_t\cm{\pitbar_t,\gbar_t} \defeq \mE^{\pitbar_t,\gbar_t}\left[R_t\cm{\Xbar_t,\Abar_t}\right.\nn\\
        \left.+ \delta \Vhat_{t+1}\cm{\Ftbar\cm{\pitbar_t, \gbar_t, \Abar_t}}\right],
      \end{align}
      where the expectation is over the random variables $\cm{\Xbar_t,\Abar_t}$ with the distribution $\gbar_t\cm{\cdot\vert\xbar_t}\pitbar_t\cm{\xbar_t}$.
      \item We define $\Ttbar\coloneqq\cm{\Thetahat_i}_{i\in\cN}$ as the estimated optimal policy which might not be the true optimal policy. $\forall \pibar \in \times_{i\in\cN} \Delta(\cX^i)$, let $\gtbar=\Ttbar_t\sq{\pitbar_t}$ be generated as follows
      \begin{subequations}
        \begin{align}
          \gtbar_t \in \arg\max_{\gbar_t} \ \Qhat_t\cm{\pitbar_t,\gbar_t} , \label{Eqn:Optim1}\\
          \Vhat_t\cm{\pitbar_t} = \Qhat_t\cm{\pitbar_t, \gtbar_t}.\label{Eqn:Vdef1}
        \end{align}
      \end{subequations}
    \end{enumerate}
  \end{enumerate}
  
  We could proceed in a similar way for the infinite horizon case $\mathfrak{D}_\infty$ where the iteration steps remain unaltered but the functions $\Vhat_t$, $\Qhat_t$ and the policy $\Ttbar_t$ become independent of time. The modified algorithm is stated as follows with $T\to\infty$.

  \begin{enumerate}
    \item Initialize $\forall \pitbar\in \times_{i\in\bN} \Delta(\cX^i)$,
    \begin{align}
      \Vhat\cm{\pitbar} \defeq 0 \label{Eqn:VT+1_2} 
    \end{align}     
    \item For $t = T,T-1, \ldots 1$,  
    \begin{enumerate}
      \item Compute $\Qhat \ \forall\pitbar\in \times_{i\in\bN} \Delta(\cX^i)$ and $\forall\gbar\in \times_{i\in\bN}\Gamma^i$ similar to the equation in~\eqref{Eqn:Bellman_Q_update}, as follows:
      \begin{align}
        \label{Eqn:Qdef2}
        \Qhat\cm{\pitbar,\gbar} \defeq \mE^{\pitbar,\gbar}\left[R\cm{\Xbar,\Abar}\right.\nn\\
        \left.+ \delta \Vhat\cm{\Ftbar\cm{\pitbar, \gbar, \Abar}}\right],
      \end{align}
      where the expectation is over the random variables $\cm{\Xbar,\Abar}$ with the distribution $\gbar\cm{\cdot\vert\xbar}\pitbar\cm{\xbar}$.
      \item We define $\Ttbar\coloneqq\cm{\Thetahat_i}_{i\in\cN}$ as the estimated optimal policy which might not be the true optimal policy. $\forall \pibar \in \times_{i\in\cN} \Delta(\cX^i)$, let $\gtbar=\Ttbar\sq{\pitbar}$ be generated as follows
      \begin{subequations}
        \begin{align}
          \gtbar \in \arg\max_{\gbar} \ \Qhat\cm{\pitbar,\gbar} , \label{Eqn:Optim2}\\
          \Vhat\cm{\pitbar} = \Qhat\cm{\pitbar, \gtbar}.\label{Eqn:Vdef2}
        \end{align}
      \end{subequations}
    \end{enumerate}
  \end{enumerate}

%% file: 4Convergence.tex
In this section, we prove the optimality of the computed policies for our decentralized multi-agent control problem. We show that the expected sum of returns is maximized by following the policy $\Tbar$. We then show that, for tha case when \pf is used to generate the updates for the belief state, we can bound the difference in the returns by upper bounding the worst case errors in estimating $Q$ and $V$ functions which goes down to zero with the increase in the number of particles used for particle filtering. The expected sum of rewards accumulated over a time $t$ till $T$ can be expressed as,
\begin{align}
	\label{Eqn:Expected_returns2}
	J_t^{\Tbar_{t:T}}\cm{\pibar_t} = \mE^{\pibar_t}[\mE^{\Tbar_t}[R\cm{\Xbar_t,\Abar_t}
	+ \delta J_{t+1}^{\Tbar_{t+1:T}}\cm{\pibar_{t+1}}\vert\Xbar_t]],
\end{align}
where $\Tbar_{t:T}$ is the true optimal policy. This could be verified by expanding the terms and using iterated expectations. The same expression could be modified when \pf is used for estimating the belief as,
\begin{align}
	\label{Eqn:Expected_returns3}
	\Jhat_t^{\Ttbar_{t:T}}\cm{\pitbar_t} = \mE^{\pitbar_t}[\mE^{\Ttbar_t}[R\cm{\Xbar_t,\Abar_t} \nn\\
	+ \delta \Jhat_{t+1}^{\Ttbar_{t+1:T}}\cm{\pitbar_{t+1}}\vert\Xbar_t]],
\end{align}
where $\Ttbar_{t:T}$ is the computed optimal policy for this case. We put forth two lemmas that show that the value function $V_t\cm{\pibar}$ in~\eqref{Eqn:Bellman_V_update1} captures the expected sum of returns till time $t$ if we follow the policy $\Tbar_{t:T}$.
\begin{lemma}
	\label{Lemma:optimal_return_is_V} 
	$\forall t\in\sq{T}$, $\forall \pibar_t\in\times_{i\in\bN}$,
	\begin{align}
		\label{Eqn:lemma1_to_prove}
		V_t\cm{\pibar_t} = \mE^{\pibar_t,\Tbar_t}\sq{R\cm{\Xbar_t,\Abar_t} + \delta J_{t+1}^{\Tbar_{t+1:T}}\cm{\pibar_{t+1}}},
	\end{align}
	where $\pibar_{t+1}=\Fbar\cm{\pibar_t,\abar_t,\gbar_t}$, $\Tbar_t\cm{\pibar_t}=\gtbar_t$ and $\cm{\Xbar_t,\Abar_t}\sim\pibar_t\cm{\xbar_t}\gtbar_t\cm{\abar_t\vert \xbar_t}$.
	
	\begin{proof}
		Refer to Appendix A.
	\end{proof}

\end{lemma}

\begin{lemma} \label{Lemma:V_is_better}
	$\forall \pibar_t\in\times_{i\in\bN}$, $\forall t\in\sq{T}$, and any policy $\boldsymbol{\theta}_t$,
	\begin{align}
		\label{Eqn:lemma2_to_prove}
		V_t\cm{\pibar_t} \geq \mE^{\pibar_t,\boldsymbol{\theta}_t}\sq{R\cm{\Xbar_t,\Abar_t} + \delta V_{t+1}\cm{\pibar_{t+1}}},
	\end{align}
	where $\gbar_t = \boldsymbol{\theta}_t\cm{\pibar_t}$, $\pibar_{t+1} = \Fbar\cm{\pibar_t,\gbar_t, \Abar_t}$ and $\cm{\Xbar_t,\Abar_t}\sim\pibar_t\cm{\xbar_t}\gbar_t\cm{\abar_t\vert \xbar_t}$.
	\begin{proof}
		This follows directly from the equations~\eqref{Eqn:Bellman_V_update1} and~\eqref{Eqn:Bellman_V_update2}.
	\end{proof}

\end{lemma}

\begin{theorem}
	\label{Thm:main_theorem}
	The optimal policy $\Tbar$ that is derived from the algorithm is the optimal strategy for the multi-agent problem.
	\begin{align}
		\label{Eqn:main_thorem}
		 J_t^{\Tbar_{t:T}}\cm{\pibar_t}\geq J_t^{\boldsymbol{\theta}_{t:T}}\cm{\pibar_t}  \ \forall \boldsymbol{\theta}.
		\end{align}

		\begin{proof}
			Refer to Appendix B.
		\end{proof}

\end{theorem}	

\subsection{Convergence with \pf}
	
	The convergence of the particle filter to the true distribution has already been established under weak conditions and law of large numbers. In this section, we upper bound the error in the estimating the returns assuming an empirical distribution instead of the true distribution in terms of the parameters of the particle filter. Convergence of $\pihat$ to $\pi$ was discussed in~\cite{douc2007limit} using law of large numbers and central limit theorems. It was quoted that the under weak conditions on the belief function, we have the consistency property that $\pitbar$ almost surely converges ti $\pibar$, therefore we have the following result in expectation \cite{moral2004feynman,Elvira2017},
	\begin{align}
		\text{if,}\quad\mE^{\pi}\sq{f\cm{x_t}} &= \int f\cm{x_t}d\pi\cm{x_t}\\
		\text{then,}\quad\mE^{\pihat}\sq{f\cm{x_t}} &= \int f\cm{x_t}d\pihat\cm{x_t}\\
													&= \lim_{K\to\infty}\frac{1}{K}\sum_{k = 1}^Kf\cm{x_t^k}.
	\end{align}
 	The use of random sampling techniques such as \glspl{pf}, where the output belief could be different for the same input belief, the same policy $\gamma$ and the observation $a$. Moreover, a slight change in the values of the belief update might lead to a bad action and thus cause a loss in optimality. Here, we intend to obtain the conditions for convergence and error bounds on the value functions with the use of particle filter.
 
	Let us consider the backward recursion algorithm that was explained in Section~\ref{sec:fhbr}. The algorithm basically consists of steps where $Q$ and $V$ values are computed. The error accumulated at each of these steps can then be stated as  
	 \begin{align}
		\ldots \circ \cm{e^V \circ e^Q}_t \circ \ldots \circ \cm{e^V \circ e^Q}_T.
	 \end{align}
	We begin the computation of gap in the functional values $Q_T$ and $\Qhat_T$ at the terminal state $T$ or a very large $T$ for infinite horizon cases and characterize the error $e^Q$. Now, consider,
	\begin{enumerate}
		\item The random variable representing the states of the agents $X^i$'s are statistically independent.
		\item Given the policy action $\gtbar$ and the state vectors $\Xbar$, the actions are deterministic. This is because the policy actions are pure, not mixed. This ensures that the only source of randomness is due to the state vectors. 
	\end{enumerate}
	 Therefore, knowing that the function of independent random variable are also independent, we can safely conclude that the rewards $R\cm{\Xbar_T,\Abar_T}$ are independent for different realizations of $\Xbar_t\sim\pibar_t$. Also let's assume the rewards are bounded within the range $r_{max}$ and $r_{min}$.
	 
	 For $N$ agents with each agent having an independent particle filter with $K$ finitely many particles, we can apply Chernoff-Hoeffding bounds on agent $i$, we get~\cite{hoeffding1994probability},
	 
	 \begin{align}
		 \mP\Big(\Big\vert\mE^{\pitbar_T,\gbar_T}\sq{R\cm{\Xbar_t,\Abar_t}}-&\mE^{\pibar_T,\gbar_T}\sq{R\cm{\Xbar_t,\Abar_t}}\Big\vert\leq\epsilon\Big)\nn\\
		 &\geq 1 - 2e^{-\frac{2K^2\epsilon^2}{K \cm{r_{max}-r_{min}}^2}}.
	 \end{align}
	 Using the definition of $Q$-value we get,
	\begin{align}
		\label{Eqn:lemma3_main}
		\mP\cm{\Big\vert \Qhat_T\cm{\pitbar_T,\gbar_T}-Q_T\cm{\pibar_T,\gbar_T}\Big\vert\leq\epsilon}&\geq 1-e^{-\frac{2K\epsilon^2}{\Delta_R^2}}\nn\\
																									&=1-\zeta,
	\end{align}
	where $\Delta_R = r_{max}-r_{min}$ and $\zeta = e^{-\frac{2K\epsilon^2}{\Delta_R^2}}$. The worst error that could occur is given as $\frac{Kr_{max}-Kr_{min}}{K}=\Delta_R$. The expected error accumulated can be expressed as 
	\begin{align}
		e^Q_T = \cm{1-\zeta}\epsilon + \zeta\Delta_R.
	\end{align}		
	Now, 
	\begin{align}
		V\cm{\pibar_T} = Q\cm{\pibar_T,\gtbar_T},
	\end{align}
	where $\gtbar_T$ is the equilibrium policy. So, $Q\cm{\pibar_T,\gtbar_T}\geq Q\cm{\pibar_T,\gtbar^\prime_T}$ for any other policy $\gtbar^\prime_T$. 
	
	From~\eqref{Eqn:lemma3_main}, we know that, with probability $1 - \zeta$, 
	\begin{align}
		\Qhat\cm{\pitbar_T,\gtbar_T} &= Q\cm{\pibar_T,\gtbar_T}-\epsilon\label{Eqn:Q1}\\
		&\leq Q\cm{\pibar_T,\gtbar^\prime_T}\label{Eqn:Q2}\\
		&\leq Q\cm{\pibar_T,\gtbar_T}\label{Eqn:Q3}\\
		&\leq Q\cm{\pibar_T,\gtbar^\prime_T}+\epsilon=\Qhat\cm{\pitbar_T,\gtbar_T^\prime}\label{Eqn:Q4}.
	\end{align}
	\eqref{Eqn:Q1}-\eqref{Eqn:Q4} gives a worst case scenario for the errors between the estimated and the true values for the value function $V$w hen the error between the $Q$ values is $\epsilon$.
	Therefore, 
	\begin{align}
		\Vhat\cm{\pitbar} = \Qhat\cm{\pitbar_T,\gtbar^\prime_T}\\
		\Big\vert\Vhat\cm{\pitbar}-V\cm{\pibar}\Big\vert = \epsilon
	\end{align}
	The worst error is $\epsilon$. Similarly, we can show in the worst case,
	\begin{align}
		\Big\vert\Vhat\cm{\pitbar}-V\cm{\pibar}\Big\vert = \Delta_R
	\end{align}
	which leads to
	\begin{align}
		\label{Eqn:eta_error}
		e^V_T = \cm{1-\zeta}\epsilon + \zeta\Delta_R \quad\quad\cm{=\eta}.
	\end{align}

	At any time $t$, in order to the estimate the error of $e^Q_t$, we use the error $e^V_{t+1}$ which estimates the error between $V_{t+1}\cm{\pibar_{t+1}}$ and $V_{t+1}\cm{\pitbar_{t+1}}$. But, 
	\begin{align}
		\pitbar_{t+1} &= \Ftbar\cm{\pitbar_t, \gtbar,\abar_t}\\
					&\neq \Ftbar\cm{\pibar_t, \gtbar,\abar_t},
	\end{align}
	where $\Ftbar\cm{\cdot}$ represents the $N$ \pf filters with $K$ particles. We assume that the particle filters are capable of estimating the future belief states from the past estimated belief states with reasonable accuracy, thus we have
	\begin{align}
		\boldsymbol{d}\cm{\boldsymbol{\mu},\boldsymbol{\nu}}\leq\beta,
	\end{align}
	where $\boldsymbol{\mu} = \Ftbar\cm{\pibar,\gtbar_t,\Abar}$ and $\boldsymbol{\nu} = \Ftbar\cm{\pitbar,\gtbar_t,\Abar}$ and $\boldsymbol{d}\cm{\cdot,\cdot}$ refers to the Wasserstein distance between two distributions~\cite{Subramanian2019}. Assuming that the value function is Lipschitz we can limit the error as
	\begin{align}
		\big\vert\Vhat\cm{\pitbar}-\Vhat\cm{\pibar}\big\vert\leq m\beta,
	\end{align}   
	where $m$ is the a Lipchitz constant for the function $\Vhat$ in the metric space for belief vector $\pitbar$~\cite{Subramanian2019}. In that case, 
	\begin{align}
		e^Q_{T-1} = \eta + \delta m\beta\\
		e^V_{T-1} = \eta + \delta m\beta,
	\end{align}
	where $\eta$ is given in~\eqref{Eqn:eta_error}. Now Considering all errors from time $t$ to $T$
	we get the accumulated error,

	\begin{align}
		\big\vert\Vhat\cm{\pitbar_t}-V\cm{\pibar_t}\big\vert\leq E,
	\end{align}   
	where 
	\begin{align}
		E&=\eta +\delta \cm{\eta + \beta}+ \ldots\nn+\delta^{T-t}\cm{\eta + \beta}\\
		&= \eta + \delta \frac{1-\delta^{T-t-1}}{1-\delta}\cm{\eta+\beta}. \label{Eqn:Convergence_error}
	\end{align}
	According to the Lemma~\ref{Lemma:optimal_return_is_V}, the returns $J_t$ is captured by the value function and thus the error in the returns is also upper bounded. Moreover, the rate of convergence of the particle filter method is independent of the dimensions of the state and the action spaces~\cite{Crisan2002,Douc2005}. Similarly, this error is bounded under the assumption that $\Delta_R$ is bounded. The statistical independence of the states cannot be guaranteed for any time instant after $t$. However, it was shown by the authors in~\cite{schmidt1995chernoff} that Chernoff-Hoeffding bounds are applicable even in cases with limited statistical independence.

%% file: 5Algorithm.tex
In this section, we describe the model-free \rl algorithm where the optimal policy is learned over a series of episodes of sampled trajectories for a specific smart grid environment. The algorithm that we use is an off-policy online \rl algorithm where different policies are used for exploration and exploitation. In addition, the set up treats the model as a separate black box whose dynamics are unknown to the algorithm and the only the reward samples corresponding to the states and actions could be viewed. We need to make an assumption that the particle filter has access to the model such that it can generate the next state samples from the samples of present state without the explicit knowledge of the transition matrix.

\subsection{Policy Module}
	The policy module implements the equation in~\eqref{Eqn:Optim2}. It takes the $\Qhat_t$ values computed in the previous episode and the estimated belief $\pitbar$ and outputs the optimal policy $\gtbar_t$. The algorithm implemented is an off policy algorithm which means, bootstrapping for the future values is done assuming a greedy policy, while the actions in the current state are determined using a sub optimal policy like the $\epsilon$-greedy policy with an appropriately chosen $\epsilon$. There are several alternate ways to implement this policy module and one of the most extensively used is based on \neural based policy gradient approach. In this paper, we tackle a simpler problem with limited number of possible policies which is why we avoid the \neural approach.

\subsection{Particle Filter}

	The issue with any \pomdp, when the model is unknown, is the computation of the updated belief state for the next instant. With no information of the transition probabilities, it is impossible to evaluate the Bayes' equation in~\eqref{Eqn:Bayes}~\cite{Subramanian2019}. Under the assumption that the underlying \mdp governing the agents evolves as conditionally independent processes, a series of particle filters, one for each agent, are employed to estimate the updated belief state $\pitbar_{t+1}$ from the current belief $\pitbar_t$. The other inputs to the modules are the corresponding $\gbar_t$ and the observed action $\abar_t$. The module runs the bootstrap filter algorithm with importance sampling and multinomial resampling as described in Section~\ref{sec:Prelims} with $K$ representing the number of particles used for estimation. 

\subsection{Policy Evaluation}

	In case of the episodic task when the horizon is time limited, the $Q$-function value is computed as function of the states as well as the time instant during each episodes. For each time instant in each of the episode, the $Q_t$ value is updated through the sarsa technique, with the target given by~\eqref{Eqn:Qdef2}, as 
	\begin{align}
		Q_t\cm{\pitbar,\gbar_t} = Q_t\cm{\pitbar,\gbar_t} + \alpha\cm{\text{Target} - Q_t\cm{\pitbar,\gbar_t}}
	\end{align}
	where $\alpha$ is the learning parameter. At each instant in the trajectory, the updated belief state given as $\pitbar_{t+1}=\hat{\Fbar}\cm{\pibar_t, \Abar_t, \gbar_t}$. The value function that was obtained from the previous episode is used to get the value corresponding to the updated belief state estimated by the particle filter $V\cm{\pitbar}$. In case of infinite horizon, the $Q$ and the $V$ functions become agnostic to the time instant and thus the updates are not to the same functions for every instant. The algorithms used for finite and infinite horizon are depicted in Alg~\ref{alg:OnlineRL1}and ALg~\ref{alg:OnlineRL2}.

\subsection{Environment}

	In order to showcase our algorithm, we implement a smartgrid problem that was specified in~\cite{subramanian2018reinforcement}. The system consists of $N$ cooperative agents with the same state and the action spaces given as $\cX=\left\{0,1\right\}$ and $\cA=\left\{\varnothing,0,1\right\}$respectively. The dynamics of the model $\mP\cm{x_{t+1}^i\vert x_t^i, a_t^i}\forall x_{t+1},x_t\in \cX$, and $a_t\in\cA$ is given as

	\begin{align}
	\mP\left\{\cdot\vert\cdot,\varnothing\right\} &= M\\ 
	\mP\left\{\cdot\vert\cdot,0\right\} &= \left(1-\epsilon_1\right)	\left(\begin{array}{cc} 1 & 0\\ 1 & 0 \end{array}\right	)  + \epsilon_1 M\\
	\mP\left\{\cdot\vert\cdot,1\right\} &= \left(1-\epsilon_2\right)	\left(\begin{array}{cc} 0 & 1\\ 0 & 1 \end{array}\right	)  + \epsilon_2 M
	\end{align}
	The reward across time $R_t$ is a given as 
	\begin{align}
	R_t=-\cm{\frac{1}{n}\sum_{i\in N}\cm{c_0\mathbbm{1}_{a_t^i=0}+c_1\mathbbm{1}_{a_t^i=1}} +KL\cm{z_t||\zeta}},
	\end{align}
	where the constants $c_0$ and $c_1$ are the costs of taking the corresponding actions. $\zeta$ represents the target distribution, $z_t$ represents the population distribution at any moment and $KL\left(z_t||\zeta\right)$ represents the Kullback-Leibler divergence between the distributions. 

	\begin{algorithm}[!htb]
		\label{alg:OnlineRL1}
		\DontPrintSemicolon
		\SetAlgoLined
		\KwIn{\\E: No. of Epsiodes\\
			T: Time Horizon\\
			$\delta$: Discount\\
			$\alpha$: Learning Parameter\\
			$\epsilon$: greedy Parameter}
		\vspace{.1cm}
		\KwOut{$\Ttbar_{t:T}$}
		\vspace{.1cm}
		Initialize $V$, $Q$\;
		\For{each episode}
		{
			Initial Belief: $\pitbar_0$\;
			Initial State: $\Xbar_0\sim\pitbar_0\cm{\cdot}$\;	
			\For{$t=0\ldots T$}
			{
				$\gbar^\epsilon_t,\gtbar_t =$ Policy Module$\cm{Q_t,\pitbar_t}$\;
				$V_t\cm{\pitbar_t} = Q_t\cm{\pitbar_t,\gtbar_t}$\;
				$\Ttbar_t\cm{\pitbar_t}=\gtbar_t$\;
				$R, \Abar_t, \Xbar_{t+1}=$ Model$\cm{\Xbar_t,\gbar_t^\epsilon}$\;
				$\pitbar_{t+1}=$Particle Filter$\cm{\pitbar_{t},\gbar_t^\epsilon,\Abar_t}$\;
				Sarsa Target: $G=R+\delta V_{t+1}\cm{\pitbar_{t+1}}$\;
				$Q$ update: $Q_t\cm{\pitbar_t,\gbar_t^\epsilon}=Q_t\cm{\pitbar_t,\gbar_t^\epsilon}+\alpha\cm{G- Q_t\cm{\pitbar_t,\gbar_t^\epsilon}}$ 
			}
		}
		\caption{Finite Horizon \rl Algorithm}
	\end{algorithm}

	\begin{algorithm}[!htb]
		\label{alg:OnlineRL2}
		\DontPrintSemicolon
		\SetAlgoLined
		\KwIn{\\T: Trajectory length\\
			$\delta$: Discount\\
			$\alpha$: Learning Parameter\\
			$\epsilon$: greedy Parameter}
		\vspace{.1cm}
		\KwOut{$\Ttbar$}
		\vspace{.1cm}
		Initialize $V$, $Q$\;
		Initial Belief: $\pitbar_0$\;
		Initial State: $\Xbar\sim\pitbar_0\cm{\cdot}$\;	
		\For{always}
		{
			$\gbar^\epsilon,\gtbar =$ Policy Module$\cm{Q,\pitbar}$\;
			$V\cm{\pitbar} = Q\cm{\pitbar,\gtbar}$\;
			$\Ttbar\cm{\pitbar}=\gtbar$\;
			$R, \Abar, \Xbar^\prime=$ Model$\cm{\Xbar,\gbar^\epsilon}$\;
			$\pitbar^\prime=$Particle Filter$\cm{\pitbar,\gbar^\epsilon,\Abar}$\;
			Sarsa Target: $G=R+\delta V\cm{\pitbar^\prime}$\;
			$Q$ update: $Q\cm{\pitbar,\gbar^\epsilon}=Q\cm{\pitbar,\gbar^\epsilon}+\alpha\cm{G- Q\cm{\pitbar,\gbar^\epsilon}}$\;
			$\Xbar=\Xbar^\prime,\pitbar=\pitbar^\prime$ 
		}
		\caption{Infinite Horizon \rl Algorithm}
	\end{algorithm}
\subsection{Simulation}

	The simulation for the finite horizon case was done for the smartgrid system with $N=2$ agents with a discretized set of belief states and a discount factor of $0.9$. The environment parameters were chosen as in~\cite{subramanian2018reinforcement}. The learning parameter $\alpha$ was chosen to be $.95$ where as the $\epsilon$ was taken to be $.2$ for the $\epsilon$ greedy algorithm. We assumed a time horizon of 10 instants for the episodic finite horizon case. We implemented the sequential decomposition algorithm with backward recursion of Section~\ref{sec:fhbr} with the model being known to the algorithm that computes the base or the optimum returns. We then implement both with model and model-free version of the online \rl algorithm discussed in Algorithms~\ref{alg:OnlineRL1} and~\ref{alg:OnlineRL2} for finite and infinite horizons respectively. We repeat the model-free version for different values of $K$ which is the number of particles used by the \pf. The whole process was averaged with $10$ runs. The model version \rl uses the equation~\eqref{Eqn:Bayes} for updating the belief state unlike the model-free version which uses the particle filter.

\subsection{Results}
\begin{figure}[!htb]
		\centering
		\includegraphics[scale=.5]{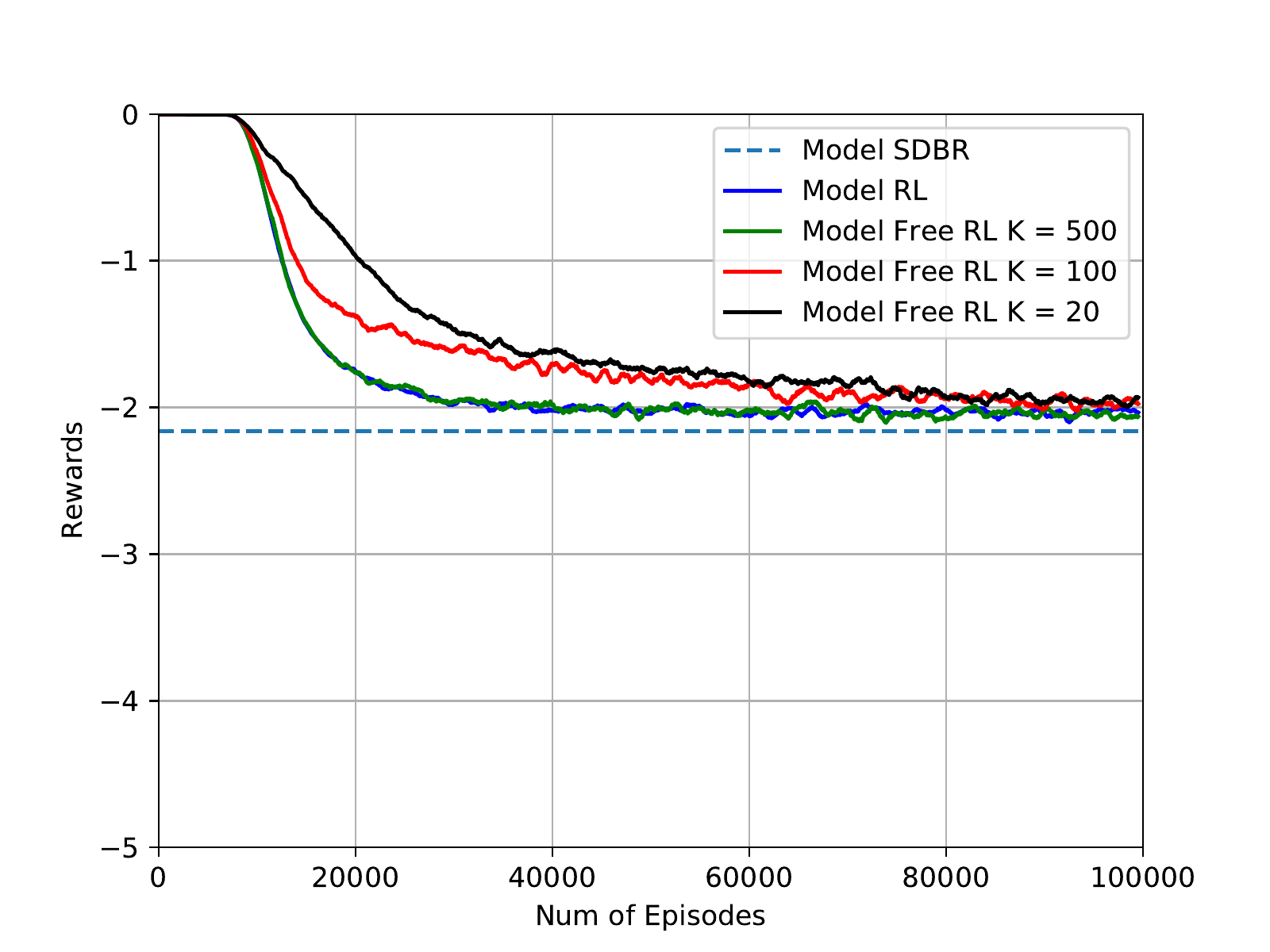}	
		\caption{Returns at the end of $10$ instants for different algorithms. The expected sum of rewards converges to the base value at different rates as a function of number of particles in the particle filter.}
		\label{Fig:plot_return_fin}
	\end{figure}

\begin{figure}[!htb]
		\centering
		\includegraphics[scale=.5]{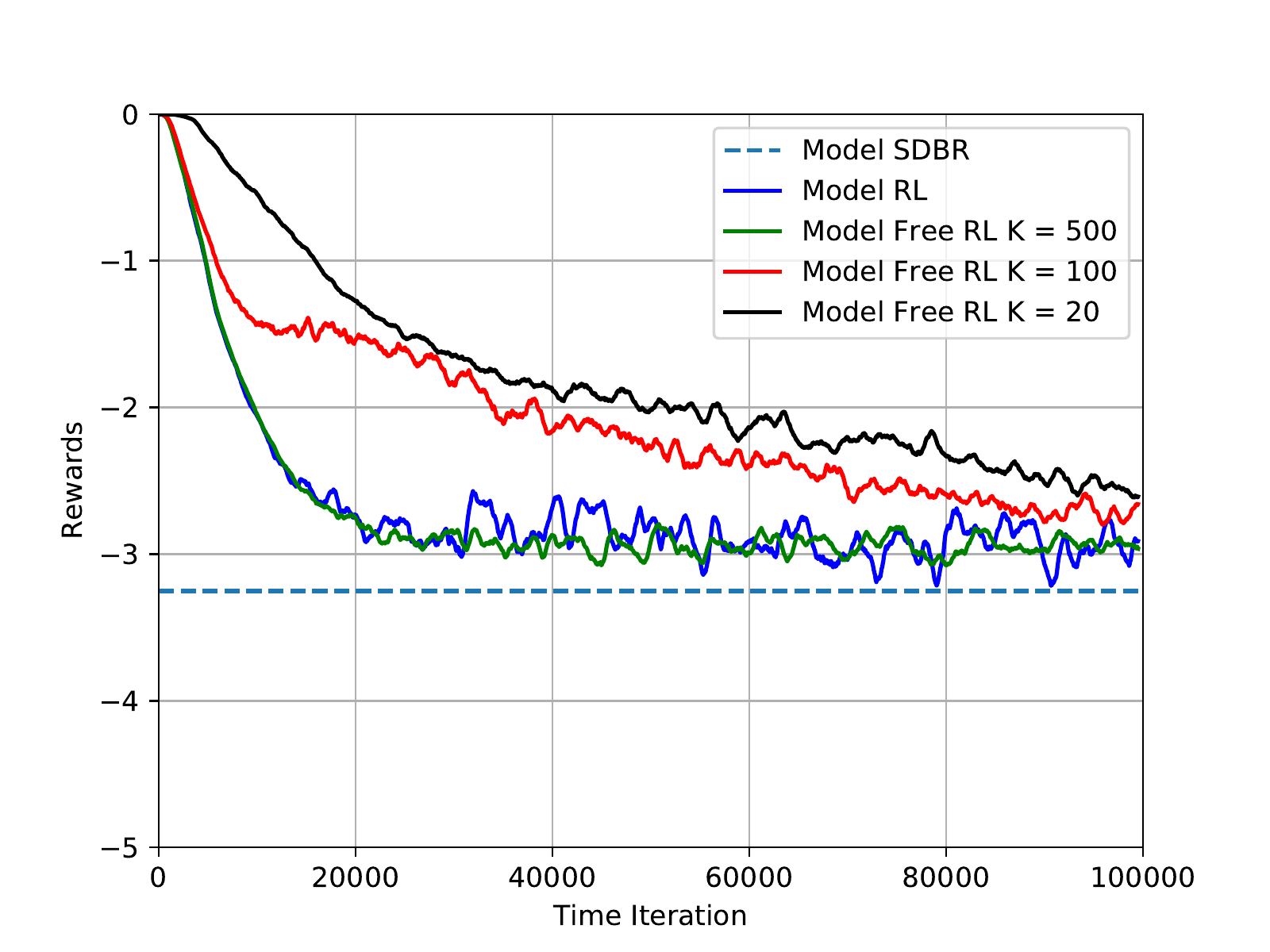}
		\caption{Returns for infinite horizon system. The returns converge to the base value at different rates. Also, the error in convergence is higher for the cases where the particle filter had less number of particles.}
		\label{Fig:plot_return_inf}
	\end{figure}

	\figref{Fig:plot_return_fin} shows the plot of the collective returns for the smartgrid system with $N$ agents with varying number of particles in the \pf. The returns shown for different algorithms are negative as they are costs incurred represented as rewards. It can be seen that the online \rl algorithm converges to the base value in all the cases though the rate of convergence is lower for lower value of $K$. The error in convergence that was shown in~\eqref{Eqn:Convergence_error} is the reason for slow rate but the error is eventually averaged out over multiple episodes which eventually makes to converge to the optimal returns. The same was again repeated assuming an infinite horizon and the plots of~\figref{Fig:plot_return_inf} was generated. Here the error difference with smaller value of $K$ is significant as it is non-episodic.

%% file: 6Conclusion.tex
In this paper, we analyzed a decentralized cooperative multi-agent system where the agents do not have access to the state information of other agents but can access the information about the actions histories of other agents. We assumed that the model of the system is unavailable and the agents need to come up with policies in order to maximize their collective rewards. We provide an \rl algorithm to compute these policies using a common agent approach wherein a fictitious common agent computes a belief on the states of the agents and then provides an optimal policy. The Bayesian updates to the belief update is impossible without the knowledge of the dynamics of the model. Therefore, we provide a \pf algorithm to compute the updated belief states and it is integrated to the \rl algorithm. We provide the proof of convergence of the model-free algorithms and the error bounds when using a \pf. We show that using \pf the optimal returns is not affected. The proposed algorithm provides a general framework to solve decentralized control problems with imperfect information and where the model information is not known. Future work would include the study of complexity analysis of the algorithm where the number of agents $N$ is large.

%% file: 7Appendix.tex
\textbf{Lemma~\ref{Lemma:optimal_return_is_V}.}
	$\forall t\in\sq{T}$, $\forall \pibar_t\in\times_{i\in\bN}$,
	\begin{align}
		V_t\cm{\pibar_t} = \mE^{\pibar_t,\Tbar_t}\sq{R\cm{\Xbar_t,\Abar_t} + \delta J_{t+1}^{\Tbar_{t+1:T}}\cm{\pibar_{t+1}}}
	\end{align}
	where $\pibar_{t+1}=\Fbar\cm{\pibar_t,\abar_t,\gbar_t}$, $\Tbar_t\cm{\pibar_t}=\gtbar_t$ and $\cm{\Xbar_t,\Abar_t}\sim\pibar_t\cm{\xbar_t}\gtbar_t\cm{\abar_t\vert \xbar_t}$.  
	\begin{proof}
		We prove the lemma using the theory of mathematical induction. 

		At $t=T$, from~\eqref{Eqn:Bellman_V_update1} and~\eqref{Eqn:Bellman_Q_update},
		\begin{subequations}
			\begin{align}
				\label{Eqn:lemma1_eqn1}
				V_T\cm{\pibar_T} =& Q_T\cm{\pibar_T,\gtbar_T}\\
				=& \mE^{\pibar_T,\gtbar_T}\sq{R\cm{\Xbar_T,\Abar_T}}.
			\end{align}
		\end{subequations}
		Since $\gtbar_t=\Tbar\cm{\pibar_t}$, we have
		\begin{align}
			V_T\cm{\pibar_T} = \mE^{\pibar_T,\Tbar_T}\sq{R\cm{\Xbar_T,\Abar_T}}.
		\end{align}
		which is the maximum returns the agents can receive at $t=T$. Now assuming that the proposition is true for $t = t+1$, we get,
		\begin{align} 	 
			V_{t+1}\cm{\pibar_{t+1}} = \mE^{\pibar_{t+1},\Tbar_{t+1}}\big [R\cm{\Xbar_{t+1},\Abar_{t+1}}+\nn\\
				\delta J_{t+2}^{\Tbar_{t+2:T}}\cm{\pibar_{t+2}}\big ] \label{Eqn:Lemma1_true_for_t+1}
		\end{align}
		At time $t = t$, we have,
		\begin{subequations}
			\begin{align}
				V_t\cm{\pibar_t} = &Q_t\cm{\pibar_t, \gtbar_t}\label{Eqn:Lemma1_eqn1}\\
									= &\mE^{\pibar_t, \Tbar_t}\sq{R\cm{\Xbar_t,\Abar_t}+\delta V_{t+1}\cm{\pibar_{t+1}}}\label{Eqn:Lemma1_eqn2}\\
									= &\mE^{\pibar_t, \Tbar_t}\big[R\cm{\Xbar_t,\Abar_t}+\nn\\
									&\delta  \mE^{\pibar_{t+1},\Tbar_{t+1}}\big[R\cm{\Xbar_{t+1},\Abar_{t+1}}+\nn\\
									&\delta J_{t+2}^{\Tbar_{t+2:T}}\cm{\pibar_{t+2}}\big]\big]\label{Eqn:Lemma1_eqn3}\\
									= &\mE^{\pibar_t, \Tbar_t}\big[R\cm{\Xbar_t,\Abar_t}+\nn\\
									&\delta  \mE^{\pibar_{t+1}}\big[\mE^{\Tbar_{t+1}}\big[R\cm{\Xbar_{t+1},\Abar_{t+1}}+\nn\\
									&\delta J_{t+2}^{\Tbar_{t+2:T}}\cm{\pibar_{t+2}}\vert \Xbar_{t+1}\big]\big]\big]\label{Eqn:Lemma1_eqn4}\\
									= &\mE^{\pibar_t, \Tbar_t}\sq{R\cm{\Xbar_t,\Abar_t}+\delta  J_{t+1}^{\Tbar_{t+1:T}}\cm{\pibar_{t+1}}}\label{Eqn:Lemma1_eqn5}
			\end{align}
		\end{subequations}			
		\eqref{Eqn:Lemma1_eqn1} is from the definition in~\eqref{Eqn:Bellman_V_update1}.~\eqref{Eqn:Lemma1_eqn2} is from the definition in~\eqref{Eqn:Bellman_Q_update}. \eqref{Eqn:Lemma1_eqn3} is from the assumption in~\eqref{Eqn:Lemma1_true_for_t+1}. Using the concepts of conditional expectation over joint random variables, we get the expression in~\eqref{Eqn:Lemma1_eqn4}.~\eqref{Eqn:Lemma1_eqn5} is from the definition of the expected sum of returns in~\eqref{Eqn:Expected_returns2}. 
	\end{proof}

\section*{Appendix B}
\textbf{Theorem~\ref{Thm:main_theorem}.}
	The optimal policy $\Tbar$ that is derived from the algorithm is the optimal strategy for the multi-agent problem.
	\begin{align}
		 J_t^{\Tbar_{t:T}}\cm{\pibar_t}\geq J_t^{\boldsymbol{\theta}_{t:T}}\cm{\pibar_t} \quad \text{for any policy} \ \boldsymbol{\theta}.
		\end{align}
	\begin{proof}
		We prove it through the technique of mathematical induction and will use the results that were proved before in Lemma~\ref{Lemma:V_is_better} and Lemma~\ref{Lemma:optimal_return_is_V}.

		For $t=T$, we need to show that,
		\begin{align}
			J_T^{\Tbar_T}\cm{\pibar_T}\geq J_T^{\Tbold_T}\cm{\pibar_T}.\label{Eqn:thm_eqn1} 
		\end{align}
		Using the expressions in~\eqref{Eqn:Expected_returns2},i.e. we can rewrite~\eqref{Eqn:thm_eqn1} as,
		\begin{align}
			\mE^{\pibar_T}&\sq{\mE^{\Tbar_T}\sq{R\cm{\Xbar_T,\Abar_T}\vert \Xbar_T}}\nn\\
			&\geq \mE^{\pibar_T}\sq{\mE^{\Tbold_T}\sq{R\cm{\Xbar_T,\Abar_T}\vert \Xbar_T}}.\label{Eqn:thm_eqn2}
		\end{align}
		Again, combining the expectations and using~\eqref{Eqn:Bellman_V_update1} we can write,
		\begin{subequations}
			\begin{align}
				\mE^{\pibar_T,\Tbar_T}\sq{R\cm{\Xbar_T,\Abar_T}}&\geq \mE^{\pibar_T,\thehat_T}\sq{R\cm{\Xbar_T,\Abar_T}}\label{Eqn:thm_eqn2}\\
				 \text{i.e.}\quad V_T\cm{\pibar_T}&\geq \mE^{\pibar_T,\thehat_T}\sq{R\cm{\Xbar_T,\Abar_T}}. \label{Eqn:thm_eqn3}
			\end{align}
		\end{subequations}
		But, we already know~\eqref{Eqn:thm_eqn3} is true as per Lemma~\ref{Lemma:V_is_better}. This proves it for $t=T$.

		Now, assuming that the condition in~\eqref{Eqn:main_thorem} holds at $t = t+1$, we get,
		\begin{align}
			\label{Eqn:Thm1_assump}
			J_{t+1}^{\Tbar_{t+1:T}}\cm{\pibar_{t+1}}\geq J_{t+1}^{\Tbold_{t+1:T}}\cm{\pibar_{t+1}}.
		\end{align}
		We need to prove that the expression in~\eqref{Eqn:main_thorem} holds for $t=t$ as well, i.e.
		\begin{align}
			J_t^{\Tbar_t}\cm{\pibar_t}\geq J_t^{\Tbold_t}\cm{\pibar_t}.
		\end{align}
		But, as per the definition in~\eqref{Eqn:Expected_returns2}, we have,
		\begin{subequations}
			\begin{align}
				J_T^{\Tbar_T}\cm{\pibar_T}=&\mE^{\pibar_t}\sq{\mE^{\Tbar_t}\sq{R\cm{X_t,A_t}+J_{t+1}^{\Tbar_{t+1:T}\cm{\pibar_{t+1}}}}}\label{Eqn:Thm1_eqn2}\\
											=&V_t\cm{\pibar_t}\label{Eqn:Thm1_eqn3}\\
											\geq &\mE^{\pibar_t}\sq{\mE^{\Tbold_t}\sq{R\cm{X_t,A_t}+J_{t+1}^{\Tbar_{t+1:T}\cm{\pibar_{t+1}}}}}\label{Eqn:Thm1_eqn4}\\
											\geq &\mE^{\pibar_t}\sq{\mE^{\Tbold_t}\sq{R\cm{X_t,A_t}+J_{t+1}^{\Tbold_{t+1:T}\cm{\pibar_{t+1}}}}}\label{Eqn:Thm1_eqn5}\\
											= &J_t^{\Tbold_t}\cm{\pibar_t}\label{Eqn:Thm1_eqn6}
			\end{align}
		\end{subequations}
			
		\eqref{Eqn:Thm1_eqn3} and~\eqref{Eqn:Thm1_eqn4} follows from the Lemmas~\ref{Lemma:optimal_return_is_V} and~\ref{Lemma:V_is_better} respectively.~\eqref{Eqn:Thm1_eqn5} is from the assumption in~\eqref{Eqn:Thm1_assump} while~\eqref{Eqn:Thm1_eqn6} follows from the definition of expected returns in~\eqref{Eqn:Expected_returns2}.
	\end{proof}